\documentclass{article}

\usepackage{amsmath}
\usepackage{arxiv}
\usepackage{tabularx}
\usepackage{array} 
\usepackage[utf8]{inputenc} 
\usepackage[T1]{fontenc}    
\usepackage{hyperref}       
\usepackage{url}            
\usepackage{booktabs}       
\usepackage{amsfonts}       
\usepackage{nicefrac}       
\usepackage{microtype}      
\usepackage{lipsum}
\usepackage{graphicx}
\usepackage{multirow}

\title{Compute-Optimal Network Design for Echocardiography Myocardial Segmentation and Perfusion Quantification using Neural Scaling Laws}

\author{
 Clara Rodrigo González \\
  Department of Bioengineering\\
  Imperial College London\\
  London, UK\\
  \texttt{cr418@ic.ac.uk} \\
   \And
 Matthieu Toulemonde \\
  Department of Bioengineering\\
  Imperial College London\\
  London, UK\\
  \texttt{m.toulemonde@imperial.ac.uk} \\
   \And
 Lasha Gvinianidze \\
  National Heart and Lung Institute\\
  Imperial College London\\
  Guy’s and St. Thomas’ NHS Foundation Trust\\
  \texttt{lasha.gvinianidze@nhs.net} \\
  \And
 Cameron A. B. Smith \\
  Department of Bioengineering\\
  Imperial College London\\
  London, UK\\
  \texttt{cameron.smith13@imperial.ac.uk} \\
  \And
 Oscar Bates \\
  Department of Bioengineering\\
  Imperial College London\\
  London, UK\\
  \texttt{o.bates18@imperial.ac.uk} \\
   \And
 Roxy Senior \\
  National Heart and Lung Institute\\
  Imperial College London\\
  Guy’s and St. Thomas’ NHS Foundation Trust \\
  \texttt{lasha.gvinianidze@nhs.net} \\
   \And
 Fu Siong Ng \\
  National Heart and Lung Institute\\
  Imperial College London\\
  London, UK\\
  \texttt{f.ng@imperial.ac.uk} \\
   \And
 Meng-Xing Tang \\
  Department of Bioengineering\\
  Imperial College London\\
  London, UK\\
  \texttt{mengxing.tang@imperial.ac.uk} \\
}

\begin{document}
\maketitle
\begin{abstract}
Myocardial perfusion quantification using contrast-enhanced ultrasound offers a bedside non-ionizing alternative to nuclear imaging modalities. However, its clinical adoption is hindered by time-consuming manual labelling. Automated segmentation has proved challenging due to a paucity of in-domain training data. Adapting strategies currently used to optimise large language models for large datasets, we apply neural scaling laws to predict network performance for myocardial segmentation. We extrapolate performance on subsets of the data to determine optimal network size on the CAMUS echocardiography dataset and a 25-patient contrast-enhanced ultrasound (CEUS) dataset. Finally, we validate the clinical utility of our models by comparing the final myocardial perfusion parameters with those obtained by a senior cardiologist. Extrapolation based on the scaling law is predictive of test loss at the full dataset size, allowing us to select two networks that obtained state-of-the-art performance on CAMUS with a 240-fold reduction in parameter count. We observe the gradient of the scaling law transfers from CAMUS to the CEUS dataset with a bias in the predicted losses. The automatically segmented masks perform equivalently to a senior cardiologist in myocardial perfusion quantification. These results establish neural scaling laws as a practical tool for data-driven compute-optimal model design for small imaging datasets.
\end{abstract}


\section{Introduction}
Myocardial perfusion imaging (MPI) is a technique used to detect abnormal blood flow into the cardiac muscle. MPI has significant diagnostic use in various conditions, including ischemic heart disease \cite{dewey_clinical_2020}, cardiac amyloidosis \cite{deux_diagnostic_2021}, and hypertrophic cardiomyopathy \cite{cao_ultrasound_2024}. MPI is performed by imaging the influx of an intravenous contrast agent into the myocardium. Traditionally, the most commonly used imaging modalities for MPI are Positron Emission Tomography (PET) and single-photon emission computed tomography (SPECT). However, these modalities expose patients to ionizing radiation; nuclear MPI has been described as the single medical test with the highest radiation burden to the US population \cite{einstein_multiple_2010}, contributing up to 20\% of the annual collective radiation dose in the United States \cite{berrington_de_gonzalez_myocardial_2010}. Only 1.5\% of imaging centers in the US met radiation dose reduction guidelines in the United States \cite{jerome_nationwide_2015}. As a result, the use of non-ionizing imaging techniques including cardiac magnetic resonance (CMR) and contrast enhanced ultrasound (CEUS) for MPI has increased significantly in recent years \cite{alskaf_deep_2022, dewey_clinical_2020}. 

Contrast-enhanced ultrasound (CEUS) uses microbubbles to achieve vascular imaging; it is non-ionizing and can be performed at the bedside. It is less costly and time-consuming than SPECT and CMR, and it is portable. CEUS has been shown to obtain comparable results to SPECT in quantifying perfusion to diagnose coronary artery disease \cite{senior_comparison_2013}. However, reliable MPI using CEUS remains a challenge due to variable image quality and artefacts, and lack of automated quantitative analysis tools \cite{dewey_clinical_2020}. Clinicians can produce qualitative assessments of myocardial perfusion by visually assessing the CEUS acquisitions. Quantitative assessment of MPI is carried out by first manually segmenting the myocardium and defining regions of interest (ROIs). The variable intensity in these ROIs can then be used to assess the influx of the contrast agent into the myocardium. However, manual labelling is time-consuming and prone to error. It works well only within frames around end systole, when there is less motion. An alternative approach using automatic segmentation would significantly decrease scan times and remove variability due to manual labelling, and it would enable the analysis of multiple myocardial areas simultaneously. 

Learning-based segmentation approaches have been applied to US images for numerous applications \cite{amiri_fine-tuning_2020,caleanu_deep_2021,chen_joint_2023}. \cite{cheng_semantic_2023} used Google's DeepLabv3+ semantic segmentation network to segment the myocardium in several views, obtaining Dice Similarity Coefficients (DSC) \cite{dice_measures_1945} around 0.86. \cite{hu_left-ventricular_2023} trained a small U-Net to segment the myocardium, the left ventricle and the left atrium in CEUS images, obtaining DSC of 0.92. While segmentation networks have been quite successful in non-contrast US, challenges remain in their application to CEUS, where intensity contrast within the chambers of the heart and the myocardium is reversed. Additionally, the limited size of available CEUS datasets constrains the achievable complexity and generalization ability of models. To our knowledge, \cite{cheng_semantic_2023} is the only work that applied these deep-learning-based segmentation to CEUS for MPI. 

In medical imaging, and CEUS in particular, access to large datasets is limited due to the cost of data acquisition. This means tuning model size as a function of the dataset is particularly important to avoid overfitting. Understanding the scaling relationship would enable improved model design and data acquisition planning.
Neural scaling analysis seeks to find a relationship between the test performance, dataset size and model size. It is mainly geared toward optimisation of large language models \cite{kaplan_scaling_2020, hoffmann_training_2022}, but some studies have also applied them to imaging related tasks, such as image reconstruction \cite{klug_scaling_2023} and text to image models \cite{li_scalability_2024}. While prior work has benchmarked model scaling empirically, comparing 18 U-Net variants across 42 medical image segmentation datasets \cite{huang_revisiting_2025}, they do not fit power laws and therefore cannot predict performance for unseen model or dataset sizes. Moreover, neural scaling requires fitting the power laws independently for each dataset, which is computationally expensive. If the fitted power laws could be applied across datasets, it would allow for better informed model design while minimizing computational cost of the scaling analysis. 

Given the limited availability of CEUS data and the added complexity of CEUS segmentation, the development of robust, generalizable learning-based models remains a challenge. Standard development requires empirically tested architecture sizes, where multiple models are trained on the complete dataset to find the best-performing one. This process is time-consuming and inefficient, often disregarding compute budgets and overlooking potentially cheaper, yet equally effective, model configurations. Therefore, there is a need for a more efficient, deterministic method to identify optimal training configurations without repeated costly retraining. 

The ability to perform MPI using ultrasound would enable faster, more accessible diagnosis for a range of cardiac conditions. This is hindered by variable image quality and the time-consuming, error-prone manual segmentation of the myocardium. In order to develop DL-based robust myocardium segmentation models with optimal model size suitable for data available, we propose the novel use of neural scaling analysis to this medical image segmentation task. Our key aims include:
\begin{itemize}
    \item To demonstrate the applicability of neural scaling analysis on small dataset scales by showing that test losses at the full dataset can be predicted by scaling analysis on smaller subsets.
    \item To explore the transferability of scaling analyses across datasets by comparing predicted and observed losses in out-of-domain datasets.  
    \item To demonstrate the clinical usefulness of these methods by applying neural scaling analysis to optimise CEUS image segmentation for automated myocardial perfusion quantification. 
\end{itemize}

\section{Related Work}

\subsection{Segmentation}
Current developments in CMR and SPECT-based perfusion quantification focus on novel methods for myocardial segmentation \cite{zhu_new_2023,entezarmahdi_qcard-nm_2023}, which reduce processing times. However, the same development is not being carried out for CEUS, where it is common to define regions of interest (ROI) manually. \cite{cao_ultrasound_2024, cosyns_how_2022} manually defined ROIs within the myocardium for the calculation of physiological parameters. 


Numerous studies have focused on the development of learning-based segmentation models for ultrasound. There has been a trend towards increasing model complexity and parameter counts to improve performance \cite{minaee_image_2022}. While the original U-Net architecture was valued for its simplicity and efficiency in small datasets, further work has focused on increasing their complexity by adding features such as attention, skip connections or transformer blocks \cite{punn_modality_2022, huang_unet_2020, azad_medical_2022}. Cardiac Acquisitions for Multi-structure Ultrasound Segmentation (CAMUS) \cite{leclerc_deep_2019} is an open-access dataset containing acquisitions from 500 patients at the University Hospital of St Etienne in France. It contains apical four-chamber and two-chamber view sequences and is commonly used for benchmarking cardiac segmentation in US. \cite{hu_left-ventricular_2023} use a U-Net with 1.9M parameters to segment the left ventricle epicardium, chamber, and atrium in CAMUS, obtaining DSC of 92\% on average across the three tissues. \cite{li_myocardial_2018} developed a statistical shape model and used a random forest to segment the pixels, and this is used to fit the shape model to the data. They obtained a DSC of 81\%. \cite{kim_automatic_2021} used a 5-layer U-Net and generative adversarial networks (GAN) to segment both the myocardium and the chamber, obtaining DSC of 0.858 with the U-Net and 0.859 with the GAN. \cite{mortada_segmentation_2023} use YOLOv7, an object detector with 37M parameters, to identify the coordinates of the left atrium and ventricle, which are then used to crop and resize the image. An 8.5M parameter U-Net is then used to segment the image, obtaining DSC of 0.856 in CAMUS. \cite{khan_compositional_2024} developed CompSeg, which uses a compositional approach and leverages metadata such as patient demographics and acquisition parameters to obtain DSC of 0.877 in the myocardium. \cite{islam_cost-unet_2024} combined convolutional and transformer blocks to segment the myocardium with DSC scores around 0.823. \cite{ullah_anatomically_2025} developed a transformer-based network and obtained DSC of 0.880 in the myocardium, 0.937 in the chamber and 0.864 in the atrium. While more advanced networks have obtained improved performance by using mechanisms such as attention or transformers, typical performance for U-Net in CAMUS obtains DSC of around 0.867 \cite{chen_daf-mamba_2025}. However, as model complexity has increased, performance has not grown at the same rate. As architectures grow in size, they encounter a performance ceiling defined by the limited diversity and scale of medical datasets, potentially leading to overfitting rather than improved generalization.  


Accurate and stable segmentation for MPQ requires the application of segmentation techniques in CEUS. This is inherently more challenging due to varying intensity \cite{carvalho_lumen_2015}, which makes the identification of tissue boundaries difficult. Additionally, the smaller datasets available for CEUS hinder the training of large networks and the comparison against baselines \cite{strohm_deep-learning-based_2025}.  \cite{chen_joint_2023} developed a transformer-based model for segmentation and diagnosis of thyroid nodules in CEUS images, obtaining DSC around 0.824. \cite{meng_ceusegnet_2022} used a  9.28M-parameter V-Net to estimate segmentation masks from paired B-Mode and CEUS images, and obtained DSC of 0.911 and IoU scores of 0.756 in a cervical lymph node dataset. \cite{li_ceus-sam_2024} created a prompt-based network based on META's Segment Anything Model (SAM) to segment breast lesions, obtaining a DSC of 0.786 and an IoU score of 0.666. \cite{duan_ai-enhanced_2026} proposed an AI-enhanced pipeline for myocardial perfusion quantification (MPQ) in myocardial contrast echocardiography (MCE), decomposing the task into segmentation, segmental division, and perfusion parameter extraction. Their segmentation model achieved a DSC of 0.88 and an IoU of 0.78 on a custom MCE dataset from Fuwai Hospital.

\label{scaling_background}
\subsection{Neural Scaling Analysis}
Neural scaling analyses seek to find the relationship between compute cost, calculated as floating point operations (FLOPs), and performance for a certain learning task. \cite{hestness_deep_2017} were the first to approach the task of empirically finding the relationship between these parameters in language tasks and image classification. They found that the test error showed a power-law relationship with dataset and model sizes. Extensive work has been done to expand these findings. \cite{rosenfeld_constructive_2019} added an extra term to the equations to capture the irreducible error, and fitted the power laws to image tasks on several image datasets, including ImageNet, CIFAR and the Describable Textures Dataset (DTD). \cite{kaplan_scaling_2020} were the first to introduce the idea of finding a trade-off between model and dataset sizes. \cite{hoffmann_training_2022} found that, for a fixed compute, it is more efficient to train smaller models on larger datasets. \cite{zhai_scaling_2022} fit these power laws to vision transformers on several datasets, including ImageNet and ObjectNet. Despite the advancements in this field, no studies were found to apply neural scaling analyses on medical images. 


 
Given a model with a total number of parameters $N$ and a dataset with $D$ images, we obtain performance loss $L(N,D)$ in the test dataset. The compute budget C is a deterministic function that depends on both $N$ and $D$. The test loss $L$ is related to the dataset size $D$ and model size $N$ by
\begin{equation}
    L(N,D) = \biggl[ \biggl( \frac{N_c}{N}\biggr)^{\frac{\alpha_N}{\alpha_D}}+\frac{D_c}{D}\biggr]^{\alpha_D}
\end{equation}

where $N_c, D_C, \alpha_N$ and $\alpha_D$ are trainable parameters. $N_c$ and $D_C$ are critical scale parameters representing the model size and dataset size, respectively, at which each term contributes equally to the total loss. $\alpha_N$ and $\alpha_D$ describe the rate at which loss decreases as model size and dataset size increase respectively, with larger values indicating faster improvement with scale. For models with a limited size, trained in large datasets, the second term of the equation can be removed. For large models trained on limited data, the first term can be removed.

When neural scaling is applied to predict test losses in a new dataset, the fitting procedure is repeated independently. This is computationally expensive and potentially redundant. We hypothesize that, if model architecture and data structure remain stable, the relative performance between the models is preserved across datasets. This would allow power laws fitted on one dataset to be applied to others. If this transferability was maintained, it would lead to scaling-informed model design without costly analysis.



\subsection{Myocardial Perfusion Quantification}
MPQ seeks to quantify several physiological parameters by monitoring the intensity-time curve within a region of interest in the myocardium. A flash-replenishment curve is drawn by plotting the acoustic intensity (AI) over time after microbubble destruction. This curve is used to estimate the proportion of tissue volume consisting of blood (MBV), the blood speed (MFV) \cite{emanuel_contrast-enhanced_2020}. These two metrics can be used to identify abnormalities in the perfusion into the myocardium. 
\section{Methods}

\subsection{Data and Preprocessing}
We used two datasets for this study. The first is Cardiac Acquisitions for Multi-structure Ultrasound Segmentation (CAMUS), an open-source dataset containing non-contrast B-mode acquisitions from 500 patients in apical four-chamber and two-chamber view sequences \cite{leclerc_deep_2019}. The dataset contains manual segmentations by three cardiologists. This dataset is heterogeneous, containing patients with pathological ejection fractions. In this study, no data selection was performed. The second dataset is a proprietary CEUS dataset acquired at Northwick Park Hospital using a Philips iE33 ultrasound machine (Philips Medical Systems, Best, Netherlands) and SonoVue (Bracco Research SA, Geneva, Switzerland) as the contrast agent. Ethical permission and informed consent were obtained, and the study was approved by the Research and Innovation department of the hospital (REC 19/LO/0624). The dataset contains acquisitions of 25 patients during rest and stress tests, and manual segmentations by a senior cardiologist. The segmentations are performed in frames around end systole, which are identified using an electrocardiogram. The dataset includes 21 healthy patients and 4 patients with coronary artery disease. Both datasets were divided into training and test sets using an 80:20 split on a patient basis. The full training/test datasets were 10,000:3,764 for CAMUS and 1,875:477 for the proprietary CEUS dataset. 

The two datasets have similar structures, with all images being acquired using diverging transmissions and the hearts in the centre of the frames. However, there are also significant differences arising from the use of contrast agents in the proprietary dataset. Variable intensity can obscure areas of the myocardium or create shadowing artefacts, making the segmentation task more challenging. 

For the MPQ, the segmentation masks were divided into six segments separating the mask in two horizontally and in three vertically. The end segments were then removed due to the lack of signal in a large proportion of the acquisitions.

\subsection{Model Architecture}
\begin{figure*}
    \centering
    \includegraphics[width=0.95\linewidth]{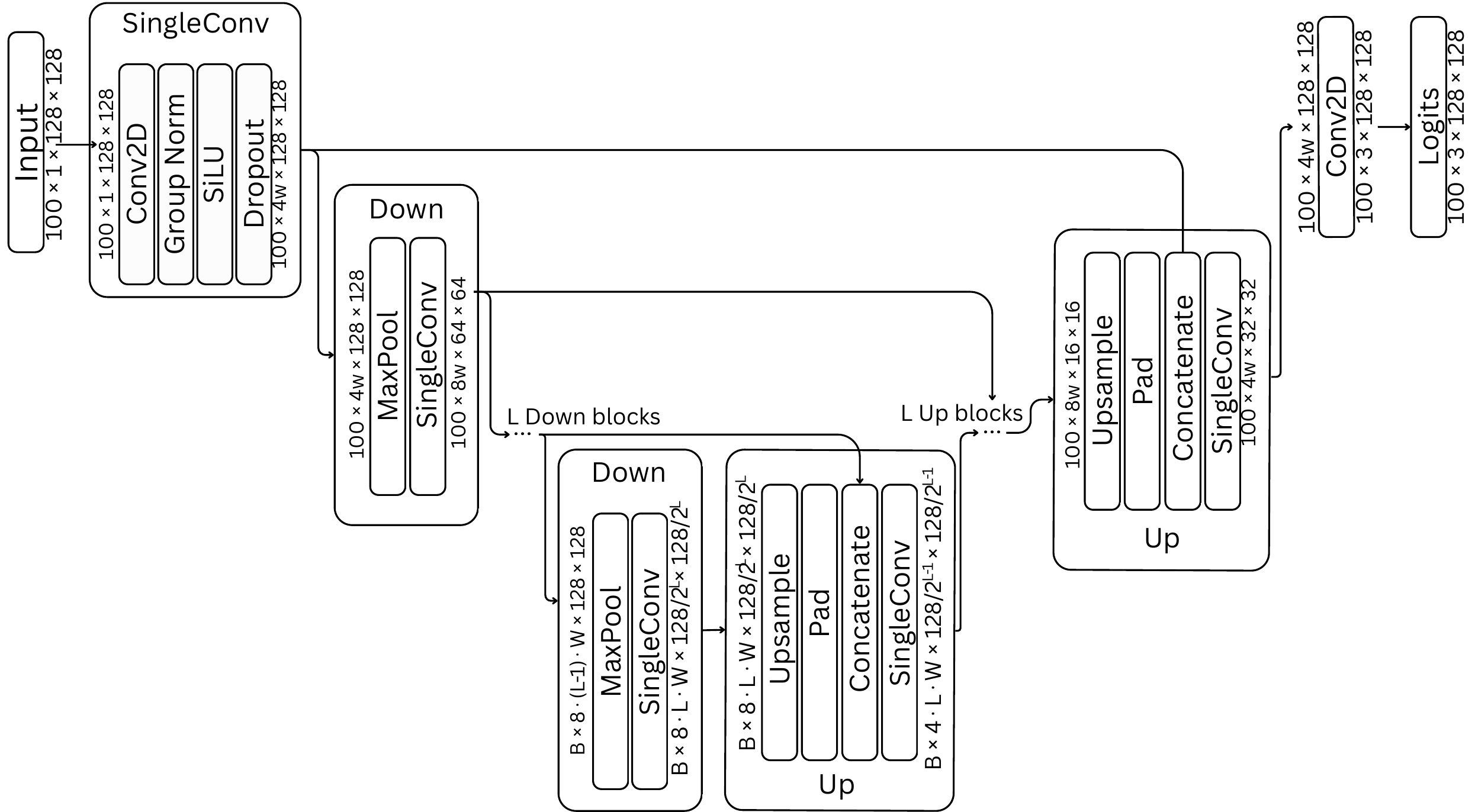}
    \caption{Our tunable U-Net is defined by parameters L, which determines the number of layers, and W, which determines the `width' of the network and how the number of channels in the data changes throughout the layers. }
    \label{fig:model_architecture}
\end{figure*}

We defined a U-Net with depth and width parameters. The U-Net receives the 1-channel image and outputs a 3-channel array with masks for the ventricle chamber, the myocardium and the atrium. Figure \ref{fig:model_architecture} shows its architecture. The depth is implemented by adding more layers, while the width is implemented by changing the number of channels of the data in the different layers. Table \ref{tab:layer_width} shows the total number of parameters for each depth-width combination. 

\begin{table}[h]
    \centering
    \begin{tabular}{|c|cccc|}
    \hline
        & \multicolumn{4}{c|}{\textbf{Width}} \\
        \cline{2-5}
        \textbf{Layers} &  \textbf{2} & \textbf{3} & \textbf{4} & \textbf{5} \\
        \hline
        \textbf{2} & & 32.9k & 58k & 90.8k \\
        \textbf{3} & 61.1k & 137k & 243k & 380k \\
        \textbf{4} & 246k & 553k & 982k & 1.5M \\
        \textbf{5} & 984k & 2.2M & 3.9M & 6.1M \\
        \hline
    \end{tabular}
    \caption{Number of parameters for different depth and width U-Nets.}
    \label{tab:layer_width}
\end{table}

The U-Nets were trained to minimize the Binary Cross Entropy (BCE) loss between their estimated masks and the ground truths. The networks were trained using an Adam optimiser with a learning rate of 0.001. The GPU used to train the models was an NVidia GeForce RTX 3090 Ti.

\subsection{Neural Scaling Analysis}
In order to fit the power law to the data, multiple models have to be trained on varying dataset sizes. The power law can then be fit to their test loss to predict performance after training on the complete dataset. 

Neural scaling laws traditionally characterise how model performance improves as a power-law function of model size, dataset size, and compute. Our approach follows the Chinchilla-style scaling law as described in Section \ref{scaling_background}, but we adapt the fitting procedure to our low-compute and smaller architecture setup. To avoid fitting all parameters simultaneously, which would be an underdetermined task with sparse data, we adopt a two-step strategy. Firstly, the data scaling exponent $\alpha_D$ is estimated globally across all architectures by fitting the full power law using the Nelder-Mead method. Since $\alpha_D$ reflects a property of the task and data distribution rather than any individual architecture, it can be held constant. In the second step, $\alpha_D$ is fixed and, for each architecture, the first term of the scaling law $( \frac{N_c}{N})^{\frac{\alpha_N}{\alpha_D}}$ is treated as a constant $A$. This decreases the number of fitting parameters to two, $A$ and $D_C$, which are fit from the BCE test loss at the inflection point obtained by training on subsets of the datasets with 500, 1000 and 5000 images. Then, the fitted curve is used to predict performance on an extended dataset size of 10,000 images in CAMUS, and 1875 in our proprietary CEUS dataset. Separate models were trained in the CEUS dataset and the prediction accuracy was evaluated. The accuracy of the fitted power laws was assessed using regression and Bland-Altman analyses. 

\subsection{Evaluation metrics} 
We evaluated the error in the segmentation masks by comparing them against ground truths using DSC, intersection over union (IoU), accuracy (ACC), precision and recall. 

To compare against a baseline from the literature, we implemented CompSeg from \cite{khan_compositional_2024}, which uses a hierarchical decoder strategy with super-segmentation and sub-segmentation components. Following their ablation study, we trained the compositional segmentation architecture without the metadata integration module (CMFI). This model has 14.7M parameters. 

The performance metrics of the models were compared using Wilcoxon signed-rank tests. We used a one-tailed Wilcoxon signed-rank test to evaluate the null hypothesis that our 61.1k parameter model is comparable to CompSeg. We used a two-tailed Wilcoxon signed-rank test to evaluate whether the larger model is more accurate than CompSeg. To correct for multiple comparisons that arise from assessing multiple metrics per model, we applied the Hommel correction ($\alpha=0.05$)  \cite{hommel_stagewise_1988}. In addition, we used the Rank Biserial Correlation (RBC) to report effect size of any statistically significant differences. A RBC magnitude value between 0.2-0.5 being small, between 0.5-0.8 being moderate and a value above 0.8 being large. 

\subsection{Myocardial Perfusion Quantification}
MPQ is carried out by monitoring the intensity curve within each segment in the myocardial mask after microbubble destruction, and fitting a saturation curve to this time-intensity curve:
 $I(t) = MBV(1 - e^{-MFV·t})$, where $MBV$ is the asymptotic value of the time-intensity curve indicating the proportion of tissue volume consisting of blood, and $MFV$ determines the slope of the influx of contrast, representing the volume of blood flowing per unit time \cite{emanuel_contrast-enhanced_2020}. 

To validate the clinical utility of our models for MPQ, we compared the MBV and MFV obtained using manually defined segmentation masks by a senior cardiologist against those obtained with two proposed U-Nets. The test set consists of 5 patients, each undergoing 10-11 acquisitions during stress and rest tests. In total, there are 57 distinct acquisitions. The perfusion parameters were aggregated to obtain one set of perfusion parameters per patient under each condition. Agreement between the perfusion parameters was reported using four metrics:
\begin{itemize}
    \item Spearman's rank correlation $\rho$ evaluates how monotonic the relationship between manual and estimated perfusion parameters is.
    \item Intraclass Correlation Coefficient (ICC) using medians, $ICC = \frac{\sigma_b^2 - \sigma_w^2}{\sigma_b^2+\sigma_w^2}$ where $\sigma_b^2=Var(\frac{y_i+\hat{y_i}}{2})$ represents the variance between methods and $\sigma_w^2=\frac{1}{2}\cdot median((y_i-\hat{y_i})^2)$ represents the variance within one method, where $y_i$ is the ground truth and $\hat{y_i}$ is the predicted value.
    \item Mean Absolute Error (MAE) between the manual and estimated parameters, defined as $MAE = \frac{1}{n}\sum_{i=1}^{n}|y_i-\hat{y_i}|$, where $y_i$ is the ground truth and $\hat{y_i}$ is the predicted value, and $n$ is the total number of values
    \item Median Absolute Percentage Error (MAPE) provides a measure of error relative to the absolute value. $MAPE = median(|\frac{y_i-\hat{y_i}}{y_i}|)×100$, where $y_i$ is the ground truth and $\hat{y_i}$ is the predicted value.
\end{itemize}

 
\section{Results}

\subsection{Neural Scaling Analysis}
The scaling analysis was carried out by fitting Equation 1 to the test performance obtained by the models on subsets of the training dataset. The parameter $\alpha_D$ was found to have a value of 0.249. We applied this fitted curve to predict the test performance on the complete dataset and assess its transferability across datasets.

\begin{figure}[!h]
    \centering
    \includegraphics[width=0.6\linewidth]{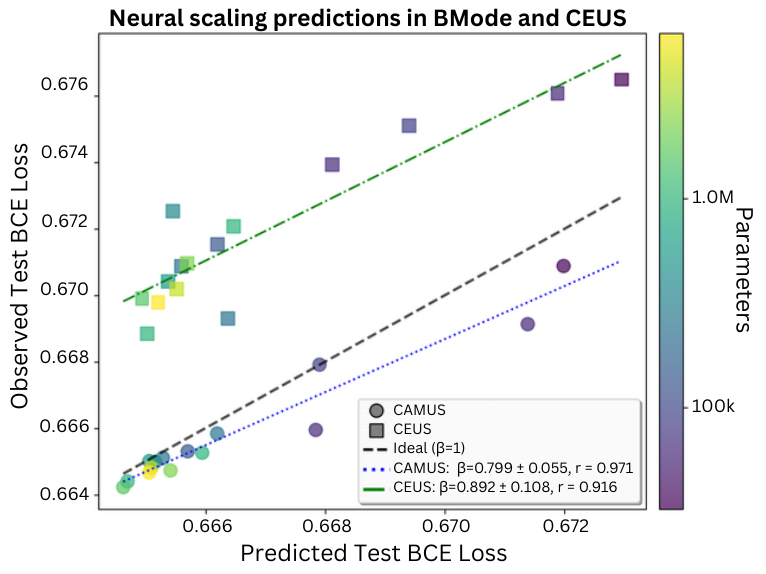}
    \caption{The test BCE at the inflection point was predicted for each model using the power law fit on models trained on subsets of the B-Mode data. These predictions are plotted against the observed losses. The dashed black line indicates perfect agreement (y=x). The blue line is the line of best fit for the CAMUS dataset, with a slope of $0.799 \pm 0.055$, and the green line is the line of best fit for the CEUS dataset, with a slope of $0.892 \pm 0.108$.}
    \label{fig:predictions}
\end{figure}

The predicted and observed losses after fitting the scaling power law on CAMUS are shown in Figure \ref{fig:predictions}. We compare the predicted losses at the 10,000-image dataset with empirically observed values (n = 15). There is a high correlation between predicted and observed values, with a Spearman correlation of $\rho$ = 0.950 (p = $6.09\cdot10^{-8}$). Linear regression between predicted and observed values yielded a slope of $0.799 \pm 0.0549$  and an intercept of $0.134 \pm 0.0366$ . While the slope deviated from unity, the high Spearman correlation indicates that the neural scaling law captured the underlying relationship between model scale and performance with high fidelity. 
  
The power law demonstrated partial cross-dataset generalization when applied to the CEUS dataset (n=15). The predicted values showed some correlation, with a Spearman of $\rho$ = 0.793  (p = $4.22\cdot10^{-4}$). Linear regression between predicted and observed values yielded a slope of $0.892 \pm 0.108$ and an intercept of $0.0768 \pm 0.0723$. 
  
A Bland-Altman analysis was performed to evaluate the agreement between the predicted and observed test losses for each dataset. For the 10,000-image CAMUS dataset, the mean difference (bias) was $-6.13\cdot10^{-4}$, confirming the slight negative systematic bias. The 95\% Limits of Agreement were found to be $-1.90\cdot10^{-4}$ to $6.78\cdot10^{-4}$. 93.3\% of points fell inside the limits of agreement. The differences were larger as the mean loss increased, suggesting improved predictive accuracy at smaller losses. A Bland Altman analysis for the CEUS dataset yielded a higher mean bias of $4.94\cdot10^{-3}$ and 95\% Limits of Agreement of $2.93\cdot10^{-3}$ to $6.95\cdot10^{-3}$. As in the CAMUS dataset, 93.3\% of points fell inside the limits of agreement.

To further assess whether the scaling laws can be applied to predict the absolute error between datasets, the data were fitted to a linear model with a shared slope and different intercepts, with equation:
\begin{equation}
    \text{observed} = \beta_0 + \beta_1 \cdot \text{prediction} + \beta_2 \cdot \text{modality}_{\text{CEUS}} + \varepsilon
\end{equation}
where $\beta_0$ is the intercept, $\beta_1$ is the prediction slope, and $\beta_2$ is the difference in intercept between CEUS and B-mode, with $\text{modality}_{\text{CEUS}}$ acting as a categorical variable. The value of $\beta_2$ was found to be 0.00562 with a $p$-value of $2.55\cdot 10^{-17}$. This $p$-value shows there is a statistically significant difference between the two modalities in their response.

\begin{figure}[!h]
    \centering
    \includegraphics[width=0.7\linewidth]{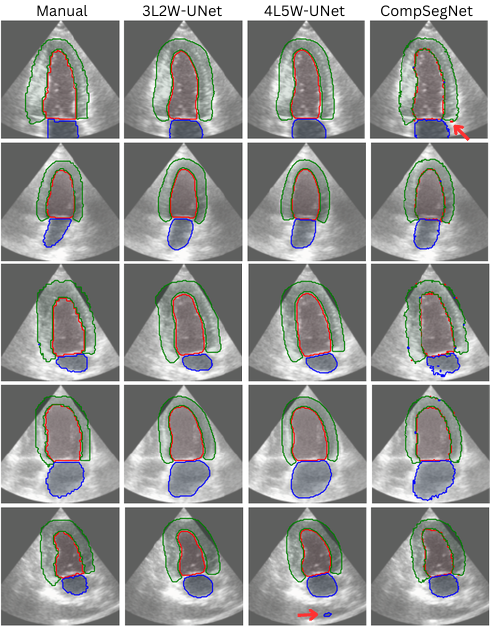}
    \caption{Example segmentations in CAMUS dataset using 3L2W-UNet and 4L5W-UNet, as well as the external SOTA model CompSeg. The underlying images are shown in grayscale, and the ventricle (red), myocardium (green) and atrium (blue) masks are shown. CompSeg created masks that resulted in some border irregularity }
    \label{fig:examples_camus}
\end{figure}

In order to validate the performance of the trained models against the SOTA from the literature, we selected a small model with 3 layers and a width of 2, and a larger 4-layer, 5-width one, which are denoted as 3L2W-UNet and 4L5W-UNet, respectively. Figure \ref{fig:examples_camus} shows example segmentations obtained with these models, as well as CompSeg. Table \ref{tab:model_comparison} shows the test performance of two chosen models in CAMUS in 3,764 images from the test set in the three tissues. 

\begin{table*}[!h]
    \renewcommand{\arraystretch}{1.2}
\newcolumntype{P}[1]{>{\centering\arraybackslash}p{#1}}
    \centering
    \footnotesize
    \begin{tabular}{|P{3cm}|P{3cm}|P{3cm}|P{3.3cm}|}
    \hline
        \textbf{Metric}  &\textbf{3L2W-UNet \newline (61.1k Parameters)} & \textbf{4L5W-UNet \newline(1.5M Parameters)} & \textbf{CompSeg \newline(14.7M Parameters)} \\
        \hline
        DSC ventricle  & 0.925 [0.905, 0.939]&\textbf{ 0.930 [0.911, 0.945]} & 0.924 [0.904, 0.937]\\
        DSC myocardium & 0.849 [0.816, 0.877]& \textbf{0.851 [0.820,0.878]} & 0.850 [0.815, 0.873]\\
        DSC atrium  & 0.894 [0.858, 0.919]& \textbf{0.900 [0.859, 0.926]}& 0.884 [0.842, 0.915]\\
        \hline
        Accuracy ventricle  & 0.964 [0.959, 0.969]& \textbf{0.965 [0.959,0.969]}& 0.963 [0.958, 0.968]\\
        Accuracy myocardium & 0.935 [0.925, 0.943]& \textbf{0.936 [0.927, 0.944]}& 0.934 [0.925, 0.943]\\
        Accuracy atrium  &\textbf{ 0.974 [0.970, 0.978]}& \textbf{0.974 [0.971, 0.977]}& 0.973 [0.969, 0.976]\\
        \hline
        IoU ventricle  & 0.763 [0.723, 0.795] & 0.778 [0.743, 0.806] & \textbf{0.780 [0.744, 0.807]}\\
        IoU myocardium & \textbf{0.663 [0.617, 0.698]}& 0.649 [0.612, 0.690] & 0.659 [0.616, 0.694]\\
        IoU atrium  & 0.687 [0.627, 0.728]& \textbf{0.706 [0.646, 0.751]} & 0.694 [0.626,  0.740]\\
        \hline
        Precision ventricle  & \textbf{0.951 [0.918, 0.972]}& 0.938 [0.900, 0.963]& 0.904 [0.859, 0.936]\\
        Precision myocardium & 0.836 [0.781, 0.875]& \textbf{0.872 [0.825, 0.910]}& 0.836 [0.788, 0.880]\\
        Precision atrium  & \textbf{0.898 [0.834, 0.950]}& 0.868 [0.799, 0.925]& 0.847 [0.777, 0.899]\\
        \hline
        Recall ventricle & 0.964 [0.929, 0.987]& 0.981 [0.952, 0.996]& \textbf{0.989 [0.968, 0.997]}\\
        Recall myocardium & \textbf{0.936 [0.895, 0.961]}& 0.912 [0.871,0.945]& 0.926 [0.890, 0.954]\\
        Recall atrium & 0.963 [0.923,0.989]& \textbf{0.989 [0.963,0.999]}& 0.987 [0.960, 0.997]\\
        \hline
         
    \end{tabular}
    \caption{Performance metrics for three models evaluated on 3,764 images in the three tissues of the CAMUS test set. Values are shown as median [IQR]. The best model for each metric and tissue is shown in bold. }
    \label{tab:model_comparison}
\end{table*}
\begin{table*}
\centering
\footnotesize
\renewcommand{\arraystretch}{1.2}
\begin{tabular}{|p{2cm}|p{2cm}|p{1cm}|p{1cm}|p{1cm}|p{1cm}|p{1cm}|}
\hline
\textbf{Model} & \textbf{Tissue} & \textbf{DSC} & \textbf{Acc.} & \textbf{IoU} & \textbf{Prec.} & \textbf{Recall} \\
\hline
\multirow{3}{*}{3L2W-UNet} & Ventricle  & +0.1336 & +0.2900 & - & +0.9618 & - \\
 & Myocardium & -0.0726 & +0.0630 & +0.0813 & - & +0.1751  \\
 & Atrium     & +0.3338 & +0.4482 & - & +0.8814 & - \\
\hline
\multirow{3}{*}{4L5W-UNet} & Ventricle  & +0.4978 & +0.5946 & +0.0916 & +0.9598 & -0.5133  \\
 & Myocardium & +0.1690 & +0.4329 & -0.2931 & +0.8888 & -0.5100  \\
 & Atrium     & +0.5987 & +0.6413 & +0.5111 & +0.6230 & +0.2496  \\
\hline
\end{tabular}
\caption{Summary of statistically significant changes relative to CompSeg in CAMUS segmentation. Values correspond to rank-biserial correlation (RBC). Positive values indicate that the reported model performed better than CompSeg, negative values indicate worse performance, and a dash (-) indicates no statistically significant difference.}
\label{tab:significance_summary}
\end{table*}

Table \ref{tab:significance_summary} shows a summary of the statistical analysis, comparing both 3L2W-UNet and 4L5W-UNet against CompSeg. On myocardial segmentation in the CAMUS dataset, the smaller 3L2W-UNet showed no significant differences against CompSeg in terms of precision. There was a slight decrease in DSC (median difference -0.0006, RBC=0.0726, p$<$0.001). The remaining metrics improved slightly, including accuracy (median difference +0.0003, RBC=0.063, p$<$0.001), IoU (median difference +0.0016, RBC=0.0813, p$<$0.001) and recall (median difference 0.009, RBC=0.175, p$<$0.001). Overall, the effect sizes are small, suggesting that the model is comparable to CompSeg, while having 240 times fewer parameters.

This behavior is maintained in the other tissues. In the ventricle, 3L2W-UNet obtains significantly improved metrics according to DSC, accuracy and precision, and obtained no significant differences in IoU and recall. The combined RBC effect size was +1.39. In atrium segmentation, 3L2W-UNet is superior to CompSeg according to the same metrics, with an overall effect size of +0.963. 

The 4L5W-UNet obtained statistically significant improvements against CompSeg across 3 of the 5 metrics studied when segmenting the myocardium. The most notable improvement was in precision (median difference +0.0355, RBC=0.889, p$<$0.001), but the model also shows improvements in DSC (median difference +0.001, RBC=0.169, p$<$0.001) and accuracy (median difference +0.0015, RBC=0.4329, p$<$0.001). The IoU decreased significantly (median difference -0.012, RBC=-0.293, p$<$0.001), and so did the recall (mean difference -0.0147, RBC=-0.510, p$<$0.001). Overall, the mean absolute effect size across all metrics was 0.688, suggesting the model provides improvements relative to CompSeg while having almost 10 times fewer parameters. 

4L5W-UNet achieved similar results in ventricular segmentation, achieving improvements in DSC, accuracy and precision, while obtaining some small decreases in IoU and recall. The overall effect size is 1.45, demonstrating significant improvements against CompSeg. 4L5W-UNet is significantly superior to CompSeg in atrial segmentation, leading to improvements in all metrics studied. This led to an overall RBC effect size of 2.62.

\begin{figure}[!h]
    \centering
    \includegraphics[width=0.9\linewidth]{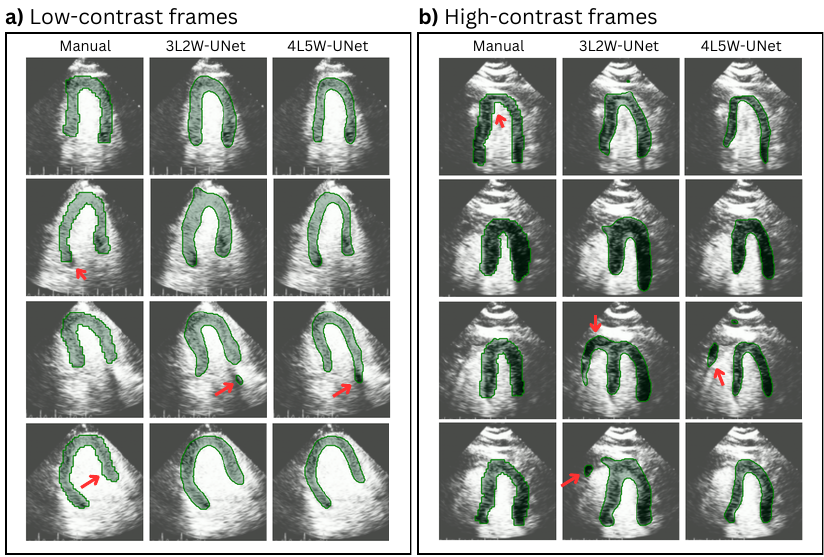}
    \caption{Randomly selected example segmentations in CEUS dataset using 3L2W-UNet (61.1k parameters) and 4L5W-UNet (1.5M parameters). The underlying images are shown in grayscale, and the masks are displayed in red. a) Low-contrast frames are shown after the contrast has had time to perfuse the myocardium, and b) High-contrast frames are extracted shortly after a high-intensity pulse destroys the bubbles. }
    \label{fig:ceus_examples}
\end{figure}

Figure \ref{fig:ceus_examples} shows example segmentations in the CEUS dataset obtained with 3L2W-UNet and 4L5W-UNet. Both networks predict masks with high agreement with the ground truth, and failure cases for both tend to be due to creating extra masks within the image or adding additional structures to the mask. 3L2W-UNet is more capable of producing these errors. Note that ground truth masks often include pixels outside the myocardium region, which could bias the perfusion parameters. Table \ref{tab:ceus_metrics} shows the performance metrics of 3L2W-UNet and 4L5W-UNet on the test set of the CEUS dataset. Both models achieve high agreement with manual labelling, with 4L5W-UNet leading to higher performance according to all metrics.

\begin{table}[!h]
    \renewcommand{\arraystretch}{1.2}
\newcolumntype{P}[1]{>{\centering\arraybackslash}p{#1}}
    \centering
    \footnotesize
    \begin{tabular}{|P{1cm}|P{3cm}|P{3cm}|}
        \hline
        \textbf{Metric} & \textbf{3L2W-UNet \newline (61.1k Parameters)}& \textbf{4L5W-UNet \newline (1.5M Parameters)}\\
        \hline
        DSC  & 0.822 [0.780, 0.852]& 0.824 [0.786, 0.855]\\
        Accuracy & 0.959 [0.947, 0.966]& 0.965 [0.956, 0.971]\\
        IoU & 0.698 [0.639, 0.742]& 0.702 [0.648, 0.746]\\
        Precision & 0.771 [0.714, 0.817]& 0.900 [0.846, 0.936]\\
        Recall & 0.895 [0.830, 0.932]& 0.775 [0.714, 0.821]\\
        \hline
        
    \end{tabular}
    \vspace{0.2cm}
    \caption{Performance metrics for two of the analysed models on 477 images in the CEUS test set. Values are shown as median [IQR]. }
    \label{tab:ceus_metrics}
    
\end{table}

\subsection{Myocardial Perfusion Quantification}
Segmentation masks obtained from the CEUS dataset were used to perform MPQ for the test patients using the two selected models, 3L2W-UNet and 4L5W-UNet. The diagnostic parameters for MBV and MFV were extracted from the fitted time-intensity curves. These were then compared to the parameters obtained after manual segmentation by a clinician. 

Table \ref{tab:agreement} shows the agreement metrics between the perfusion parameters obtained with the two models and manual labelling. Both models obtain high performance in both Stress and Rest acquisitions. The Spearman correlation coefficients ($\rho>0.87$) and high ICC (ICC$>$0.95) indicate monotonic relationships between the values obtained by both models and the ground truths. 

The models exhibited higher correlation coefficients and lower error metrics in the stress acquisitions. For example, 3L2W-UNet obtained a MAPE of 5.5\% for MFV under stress conditions, compared to 14.7\% at rest. This behavior may be attributable to the constant duration of the acquisitions in conjunction with the increased heart rate during stress. As a result, the networks were exposed to and trained on a greater number of cardiac cycles under stress than under resting conditions.

Estimations of the MBV consistently obtained lower error rates than MFV. This is expected, as MBV is calculated from the plateau of the time-intensity curve, which is stable, while MFV must be estimated from the rate of intensity increase, which is more sensitive to noise and frame rate, and curve-fitting instability in the ascending phase.

Despite a nearly 25-fold difference in their sizes, 3L2W-UNet performs comparably to the larger 4L5W-UNet. In both Stress and Rest MPQ, the smaller model obtains a comparable correlation according to Spearman and ICC metrics, and in many cases it outperforms the larger network. While both networks have comparable biases, 4L5W-UNet tended to show wider limits of agreement, suggesting that while its mean bias is comparable, its predictions are less consistent across patients.

The findings obtained from performing MPQ using 3L2W-UNet and 4L5W-UNet are consistent with the scaling law analysis, which identified the smaller network as operating near the performance saturation point for this dataset size, suggesting that additional parameters yield diminishing returns in this data-limited regime.



\begin{table*}[!ht]
\footnotesize
\renewcommand{\arraystretch}{1.2}
\newcolumntype{Y}{>{\centering\arraybackslash}X} 
\centering
\begin{tabularx}{\textwidth}{|l|Y|Y|Y|Y|}

\hline
\multirow{3}{*}{\textbf{Metric}} & \multicolumn{2}{c|}{\textbf{MBV}} & \multicolumn{2}{c|}{\textbf{MFV}} \\ \cline{2-5} 
 & \textbf{Stress} & \textbf{Rest} & \textbf{Stress} & \textbf{Rest} \\ 
 & \textit{(n=30)} & \textit{(n=27)} & \textit{(n=30)} & \textit{(n=27)} \\ \hline
\multicolumn{5}{|c|}{\textbf{3L2W-UNet (61.1k Parameters)}} \\ \hline
Spearman & 0.975 & 0.950 & 0.926 & 0.889 \\ \hline
ICC & 0.982 & 0.988 & 0.989 & 0.986 \\ \hline
MAE & 0.014 & 0.028 & 1.874 & 2.497 \\ \hline
MAPE (\%) & 2.179 [1.46, 3.00] & 2.689 [1.34, 7.05] & 5.518 [4.07, 14.86] & 14.698 [6.54, 25.36] \\ \hline
Bias & -0.008 [-0.015, 0.011] & 0.016 [-0.008,0.022] & -0.102 [-0.966, 0.367] & -0.982 [-1.970, 0.110] \\ \hline
\multicolumn{5}{|c|}{\textbf{4L5W-UNet (1.5M Parameters)}} \\ \hline
Spearman & 0.977 & 0.945 & 0.954 & 0.875 \\ \hline
ICC & 0.983 & 0.958 & 0.984 & 0.986 \\ \hline
MAE & 0.014 & 0.019 & 2.122 & 3.404 \\ \hline
MAPE (\%) & 2.077 [0.77, 3.71] & 3.254 [1.63, 3.91] & 8.761 [4.95, 14.68] & 12.76 [7.42, 34.09] \\ \hline
Bias & -0.004 [-0.015, 0.013] & 0.010 [-0.005, 0.022] & 0.280 [-0.727, 0.879] & -0.392 [-2.702, 1.044] \\ \hline
\end{tabularx}
\caption{Agreement metrics for MBV and MFV estimates across Stress and Rest acquisitions for both UNet architectures against manual labelling. Where applicable, values are shown as median [IQR]. }
\label{tab:agreement}
\end{table*}


\section{Discussion}

Neural scaling provides a systematic framework for predicting performance on the test set as a function of dataset size and network architecture, enabling improved training strategy design and achieving compute-optimal models. In this study, we demonstrated that a scaling law fitted on progressively larger subsets of the training dataset can be used to predict performance on the full dataset, allowing us to achieve SOTA performance with a significantly reduced compute budget. The relative performance between models can be predicted by transferring power laws across datasets. Finally, we showed that networks trained using this approach can be used to perform myocardial perfusion quantification with agreement comparable to manual labelling. 

The power law was fitted to the model performance at different subsets of the data, with 500, 1000 and 5000 images. This curve was then used to predict the performance of the models for the full dataset of 10,000 images. Models below a minimum capacity threshold were excluded from fitting, as they exhibited no dependence on dataset size due to their limited capacity. The slope of $0.799\pm0.055$ and the intercept of $0.0768\pm0.0723$ showed that the power law cannot be used directly to predict absolute values for each model, but it can be used to predict relative performance between models. The predictions are particularly accurate as the loss decreases, which is the primary area of focus in model development. 

When the fitted power law was applied to the CEUS dataset, relative model performance remained predictable, with a slope of $0.892 \pm 0.108$. There was a larger systematic bias in absolute loss values, which is attributable to the increased complexity of CEUS segmentation due to the decreased signal-to-noise ratio, time-varying intensity, and contrast artifacts. Notably, the slope of the line of best fit remained consistent with that of the B-Mode dataset, suggesting it is more dependent on the family of datasets rather than the data domain. This implies that a scaling law fitted on a simpler or more data-rich domain could be used to guide architecture selection in a related but more challenging domain, without requiring full retraining across all dataset sizes.

Evaluation of two of our trained models against CompSeg, a previously published architecture, showed that significant decreases in computational cost can be achieved without sacrificing performance. 3L2W-UNet (61.1k parameters) obtained a performance comparable to CompSeg with 240 times fewer parameters, and 4L5W-UNet (1.5M parameters) obtained improvements in performance with almost 10 times fewer parameters. The improvement in performance when using smaller models suggests CompSeg overfits the training dataset due to its large size. This highlights the usefulness of neural scaling laws in identifying architectures with optimal trade-offs between performance and cost. The fact that considerably smaller networks can perform comparably to larger ones in this task suggests that data-limited scenarios require fundamentally different model design strategies, and that many contemporary approaches may be unnecessarily complex.

We applied our trained models to the task of MPQ. Both networks obtained high agreement metrics against the manually obtained perfusion parameters for both MBV and MFV, demonstrating that the masks are sufficiently accurate to derive clinically relevant parameters with strong agreement to manual measurements. Performance was slightly decreased in Rest acquisitions, likely due to a slight class imbalance in the training dataset. 

While our two-step fitting approach handles limited model sizes, several limitations remain. First, the small number of dataset size increments (500, 1000, 5000 images) makes the power-law fit sensitive to noise in the BCE loss at any single point. Although the Nelder-Mead method provided a stable global $\alpha_D$, more granular increments would further refine the transition point where models hit their capacity ceilings. Additionally, the different bias between the CAMUS and CEUS datasets shows some domain-specificity, which remains unexplored. When the models were applied to CEUS segmentation, they generated artifacts in a minority of the images, such as smaller masks outside the myocardium. This could be mitigated by adding regularisation terms in the training loss. However, we kept the same training loss as in the CAMUS dataset to ensure consistency for the scaling law analysis. This highlights a potential limitation of this approach, as it constrains the extent to which hyperparameters can be adapted to dataset-specific characteristics.

We have demonstrated how neural scaling laws can be used in data-limited regimes to guide model design. We show that even with sparse dataset sizes, our two-step fitting procedure can be used to extrapolate model performance. This methodology allowed us to train a network which could match the performance of the state-of-the-art CompSeg model with a 240-fold reduction in parameters. By applying these optimized models to MPQ, we obtained clinically meaningful perfusion parameters, obtaining high agreement with manual labelling. 

\section{Conclusion}
In conclusion, this study demonstrates the utility of leveraging neural scaling analysis to guide model design. By fitting the Chinchilla power law to subsets of the training dataset, we showed that the test performance can be predicted to guide the choice of compute-optimal architectures. We applied this strategy to train 3L2W-UNet, which matched a previously published SOTA model with 240 times fewer parameters, and 4L5W-UNet, which outperformed the SOTA model with 10 times fewer parameters. We assessed the transferability of power laws across domains, showing that they can be used to predict relative but not absolute performance of the models. This indicates costly domain-specific scaling analysis is not needed. Finally, we demonstrated that trained networks can be used to derive clinically relevant perfusion parameters and obtain excellent agreement with manually derived ones. These findings establish neural scaling as an indispensable tool for designing compute-optimal and clinically robust deep learning solutions in medical imaging. 

\bibliography{references}
\end{document}